\documentstyle[12pt]{article}
\input epsf
\begin{document}

\title{An algorithm for series expansions
based on hierarchical rate equations}

\author{Chee Kwan Gan$^{a,b}$ and Jian-Sheng Wang$^a$\\
 \\
$^a$Computational Science Programme,\\
National University of Singapore, Singapore 119260\\
 \\
$^b$The Institute of Physical and Chemical Research (RIKEN), \\
Wako-shi, Saitama 351-01, Japan}

\maketitle

\begin{abstract}
We propose a computational method to obtain series expansions in powers
of time for
general dynamical systems described by a set of hierarchical rate
equations.  The method is generally applicable to problems in
both equilibrium and nonequilibrium statistical mechanics such as random
sequential adsorption, diffusion-reaction dynamics, and Ising
dynamics.  New result of random sequential adsorption of dimers on
a square lattice is presented.

\end{abstract}

Most dynamical models in equilibrium and nonequilibrium statistical
mechanics can be described by a set of hierarchical rate equations:
random sequential adsorptions (RSA) and their variants
\cite{ref:EvansReviewRSA}, diffusion-reaction
models~\cite{ref:diffusionreaction},
and kinetic Ising models \cite{ref:Glauber}.
The specifications of interactions between the components of the
system or interactions between the environment and the system give
a deterministic time evolution for the distribution functions once the
initial condition of the system is given.  Exact solutions are often
restricted to simple models, while one has to use approximate methods
for those that resist exact analyses.

Power series expansion is one of the methods of controlled
approximations.  There are a number of systematic methods to obtain
time-power series
\cite{ref:Evans1984,ref:Poland,ref:BaramKutasov,ref:DickmanVirial}.
Long series were usually obtained by reducing a problem to an
enumeration problem similar to the series expansions in equilibrium
statistical mechanics \cite{ref:Martin}.

Lattice enumeration problems are typically limited by CPU time due to
their exponential growth.  In this Letter, we discuss a general method
to obtain series expansions based on rate equations
for lattice models.  Our
method differs from previous approaches in that the limiting factor is
memory space, but it is faster in time.  The method is applicable to
many different problems, in particular RSA, RSA with diffusional
relaxation, reaction-diffusion problems, and general Ising dynamics.

To illustrate the method, we take the RSA of random dimer filling on
a square lattice as an example \cite{ref:evanstrunc}.
Dimers of random orientations are
dropped randomly and sequentially, at a rate of $k$ per lattice
site per unit time,
onto an initially empty, infinite square lattice.
Hereafter we set $k$ equal to unity without loss of generality.
If the chosen two nearest neighbor sites
are unoccupied, the dimer is adsorbed on the lattice.
If one of the chosen sites is occupied by a previously adsorbed dimer,
the adsorption attempt is rejected.
The first few of an infinite number of rate
equations for this process are given as follows:
\unitlength=6pt
\def\ck{\circle{0.6}}
\def\oA{\begin{picture}(1,1)(-0.5,-0.5)  
\put(0,0){\ck}
\end{picture}}
\def\oB{\begin{picture}(2,1)(-0.5,-0.5)  
\put(0,0){\ck}
\put(1,0){\ck}
\end{picture}}
\def\oC{\begin{picture}(3,1)(-0.5,-0.5)  
\put(0,0){\ck}
\put(1,0){\ck}
\put(2,0){\ck}
\end{picture}}
\def\oD{\begin{picture}(2,2)(-0.5,0)  
\put(0,0){\ck}                           
\put(1,0){\ck}
\put(0,1){\ck}
\end{picture}}
\def\oE{\begin{picture}(4,1)(-0.5,-0.5)  
\put(0,0){\ck}
\put(1,0){\ck}
\put(2,0){\ck}
\put(3,0){\ck}
\end{picture}}
\def\oF{\begin{picture}(3,2)(-0.5,0)  
\put(0,0){\ck}                           
\put(1,0){\ck}
\put(2,0){\ck}
\put(0,1){\ck}
\end{picture}}
\def\oG{\begin{picture}(3,2)(-0.5,0)  
\put(0,0){\ck}                           
\put(1,0){\ck}
\put(2,0){\ck}
\put(1,1){\ck}
\end{picture}}
\def\oH{\begin{picture}(2,2)(-0.5,0)  
\put(0,0){\ck}                           
\put(1,0){\ck}
\put(0,1){\ck}
\put(1,1){\ck}
\end{picture}}
\def\oI{\begin{picture}(3,2)(-0.5,0)  
\put(0,0){\ck}                           
\put(1,0){\ck}
\put(1,1){\ck}
\put(2,1){\ck}
\end{picture}}

\begin{eqnarray}
{dP(\oA)\over dt} &=& - 4 P(\oB), \label{eq:A}\\
{dP(\oB)\over dt} &=& - P(\oB) - 2 P(\oC) - 4 P(\oD), \label{eq:B}\\
{dP(\oC)\over dt} &=& - 2 P(\oC) - 2 P(\oE) - 4 P(\oF) - 2 P(\oG),
\label{eq:C}\\
{dP(\oD)\over dt} &=& - 2 P(\oD) - 2 P(\oH) - 2 P(\oG) - 2 P(\oF) - 2P(\oI),
\label{eq:D}\\
\cdots \nonumber
\end{eqnarray}
where $P(C)$ stands for the probability of finding a configuration
$C$ of sites specified empty or filled. An $\oA$ denotes an empty site
configuration.  We have taken into account
the symmetries of the problem, i.e. the invariance of $P(C)$ under
all lattice group operations (e.g. rotation and reflection). For
notational convenience we will use $P_C \equiv P(C)$.

Let $C_0$ denotes a particular configuration of interest, then
$P_{C_0}$ is the configuration probability associated with $C_0$.  On
physical grounds, we expect $P_{C_0}$ to be a well behaved function of
time $t$, and one would expect to obtain the Taylor expansion
$P_{C_0}(t) =\sum\limits_{n=0}^{\infty}{P_{C_0}}^{(n)}t^n/n!$,
with $n$-th {\it derivative} of $P_{C_0}$ given by
\begin{equation}
{P_{C_0}}^{(n)} = \left. {d^n P_{C_0}(t) \over d t^n}\right|_{t=0}.
\end{equation}

The zeroth derivative of $P_{C_0}$ is determined by the initial condition.
The first derivative of $P_{C_0}$ is obtained by
the rate equation associated with
$C_0$.  For the random dimer deposition problem, we choose
the configuration of an empty site to be $C_0$.
We have, from Eq.~(\ref{eq:A}) to (\ref{eq:D}), $P(\oA)^{(0)}
= 1$ and $P(\oA)^{(1)} = -4$, $P(\oB)^{(1)} = -7$, $P(\oC)^{(1)} =
-10$, and $P(\oD)^{(1)} = - 10$.  To compute the second derivative of
$P(\oA)$, we take the first derivative of Eq.~(\ref{eq:A}) and use the
result of $P(\oB)^{(1)}$ to obtain $P(\oA)^{(2)} = (-4)(-7) = 28$.
For the third derivative $P(\oA)^{(3)}$, we take the second derivative
of Eq.~(\ref{eq:A}), which in turn needs first derivative of
Eq.~(\ref{eq:B}). Using Eq.~(2) to (4) for the first derivatives, we
get $P(\oA)^{(3)} = - 268$.

To computerize the calculations with high efficiency, we make the
following important observations. For any configuration $C$, the rate
equation is always of the form
\begin{equation}
{d P_{C} \over dt} = \sum_{C'}\lambda_{C'} P_{C'}. \label{eq:1st}
\end{equation}
This immediately gives us the first derivative of $P_{C}$; in
particular the first derivative of $P_{C_0}$ of interest.  Since the
derivative operator is linear, the second derivative of $P_C$ is a
linear combination of the first derivatives of $P_{C'}$ on
the right-hand side of the rate equation (\ref{eq:1st}).
For each higher derivative,
new rate equations and new configurations are involved.  Let $G_i$
denote the set of new configurations generated in
the calculation of the $i$-th
derivative of $P_{C_0}$, and $G_i^j$ the corresponding $j$-th
derivatives of the set of configurations.  We observe that
$G_0^{n-1}$, $G_1^{n-2}$, $\ldots$, $G_{n-1}^0$ (determined at the
$(n-1)$-th derivative), $G_0^{n-2}$, $G_1^{n-3}$, $\ldots$,
$G_{n-2}^0$ (determined at the $(n-2)$-th derivative), $\ldots$,
$G_0^0$ are known before calculating the $n$-th derivative. In other
words, $G_i^j$ are predetermined where $ 0 \le i+j \le n-1$ at this
stage.  The derivatives $G_0^n$, $G_1^{n-1}$, $\ldots$, $G_{n-1}^1$,
$G_{n}^0$, would then be determined in a {\it bottom-up} fashion by
recursive use of the rate equations.  The method is efficient in the
sense that each value in $G_i^{n-i}$, $ 0 \le i \le n$, is calculated
exactly once and the rate equation for a configuration $C$ is also generated
exactly once.  By repeated use of this procedure, we can in
principle calculate $G_0^n$ for any $n \ge 1$.

In most cases, the knowledge of the initial condition of the system
(usually one starts with an initially empty lattice) can further
improve the efficiency of the method.  For example, let $h$ be the
highest order of derivative of $P_{C_0}$ for which we wish to compute.
Then only a subset ${G_i}',\ i \le h$, of $G_i$ needs to be considered, once
it is known that an element in $G_i - {G_i}'$ gives a zero
contribution in the subsequent calculations.  This observation is
particularly useful for models which {\it introduce} particles in
the rate equations; this is typical in models involving desorption or
diffusion.

To implement the above scheme, we store symbolically the rate
equations and the intermediate derivatives $G_i^j$.  For each
configuration $C$, we store the configuration and its first $j$
derivatives known so far.  The rate equation for $C$ is generated when
the first derivative of $P_C$ is needed, and the equation is
represented as an array of pointers to other configurations
corresponding to the right-hand side of Eq.~(\ref{eq:1st}), and the
corresponding array of coefficients $\lambda_{C'}$.

The structural part implementing the recursive use of rate equations
is independent of the details of a model.  The function calls for a
core routine to generate a rate equation if it is not available.
Symmetries of the problem are treated here.  This routine is called only
once for each configuration~$C$.
To facilitate this, only a pointer of
reference is used for each unique configuration.
Configurations are searched with the help of a hash
table or a tree data structure.

The merit of the approach allows us to obtain a series
up to $t^{14}$
for the RSA of dimers on a square lattice (Table~\ref{tab:dimer}).
Results for other models will be published elsewhere.  The
computational costs are presented
in Table 2.  Poland obtained the
same series up to $t^7$ \cite{ref:Polandseries}. For the RSA of
monomer filling with nearest-neighbor exclusion on a
square lattice, the efficiency of
our algorithm is comparable to the algorithm based on the operator
formalism \cite{ref:DickmanVirial}, but it is not as good as the method
used by Baram and Fixman ~\cite{ref:BaramFixman}.
However, the power of our method is its generality.

\begin{figure}
\caption{Pad\'e approximant estimates for the jamming coverage
$\theta_\infty$ as a function of the transformation parameter $b$,
in the crossing region.}
\label{fig:padedimer}
$$\epsfysize=9cm\hbox{\epsffile{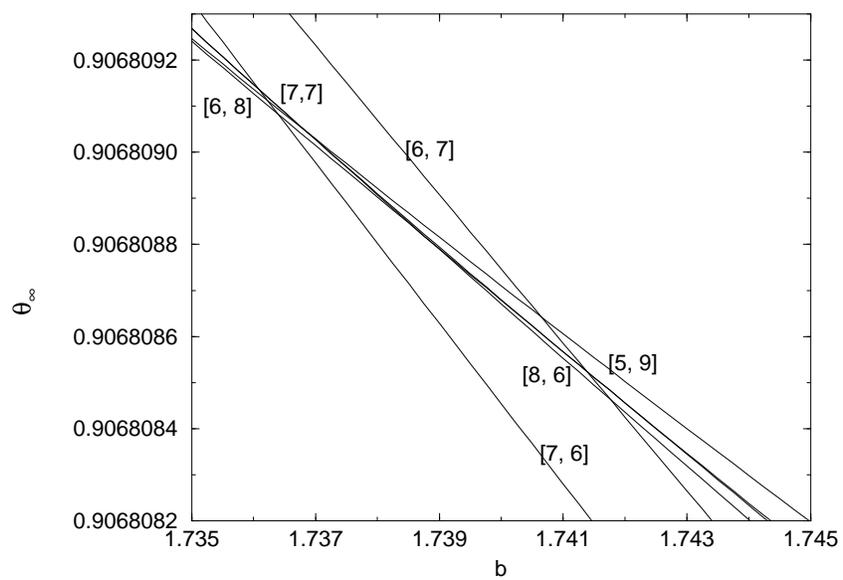}}$$
\end{figure}

The approach to the jamming state for lattice RSA is often
exponential.  We follow the analysis of series by Dickman, Wang, and
Jensen \cite{ref:DickmanVirial}.  First we transform the coverage
$\theta(t) = 1 - P(\oA,t)$ into a function of $y = 1 - e^{-t}$.
Another transformation via $z = (1-e^{-by})/b$,
is performed for the second time,
and we examine various Pad\'e approximants to the $z$
series.  As we see from Fig.~\ref{fig:padedimer}, the resulting
estimates for $\theta_\infty$ are excellent for $1.736 < b < 1.742$.
The intersections of [6, 8], [7, 7], [7, 6], [8, 6], [6, 7],
and [5, 9] approximants around this range of $b$ yield
an estimate of $\theta_\infty = 0.9068088(4)$, where the last digit
denotes the uncertainty.  This result is in good agreement with the
simulation result of $0.906873\pm 0.000138$ \cite{ref:nordsimul} or
the result of $0.9068$ obtained via approximate
truncation procedures \cite{ref:evanstrunc}.

In summary, we introduced a new approach to deriving power series
expansions in time, and we have applied the method to a RSA problem.
For the RSA of dimers on a square lattice,
we are able to generate a 15-term series,
thence to derive the most precise estimate for the jamming
coverage yet presented. Our computational method is general
and it can be used to handle a variety of problems
based on rate equations, especially
those deal with the kinetics of the lattice models.

This work is supported in part by a National University of Singapore
Academic Research Grant RP950601. Calculations were performed on the
facilities of the Computation Center of the Institute of Physical and
Chemical Research.

\newpage

\bibliographystyle{plain}

\newpage

\begin{table}
\caption{
Taylor expansion coefficients for the probability of finding an empty
site of random dimer filling on a square lattice.}
\label{tab:dimer}
\medskip
$$\vbox{\offinterlineskip\halign{
	\strut#&		
	\quad\hfil#\hfil\quad&   	
	\quad\hfil$#$ \quad&
	\quad\hfil# \quad\cr
\noalign{\hrule}\cr
& $n$ & P(\oA)^{(n)} \cr
\noalign{\smallskip\hrule} \cr
& 0 & 1 \cr
& 1 & -4 \cr
& 2 & 28 \cr
& 3 &  -268 \cr
& 4 &  3212 \cr
& 5 & -45868 \cr
& 6 & 756364 \cr
& 7 &  -14094572 \cr
& 8 & 292140492 \cr
& 9 & -6653993260 \cr
& 10 & 164952149516 \cr
& 11 & -4416119044972 \cr
& 12 & 126863203272268 \cr
& 13 & -3889473277203116  \cr
& 14 & 126677386324657804 \cr
\noalign{\smallskip\hrule} \cr
}}$$
\end{table}

\newpage
\begin{table}
\caption{CPU time and memory usage on a DEC3000/900
for the dimer problem.}
\label{tab:cpu}
\medskip
$$\vbox{\offinterlineskip\halign{
	\strut#&		
	\quad\hfil#\hfil\quad&   	
	\quad\hfil# \quad&
	\quad\hfil# \quad\cr
\noalign{\hrule}\cr
& order & memory (megabytes)& time (seconds) \cr
\noalign{\smallskip\hrule} \cr
& 9 &  1 & 1.0 \cr
& 10 & 3 & 4.4 \cr
& 11 &  12 & 22.0 \cr
& 12 & 45 & 114.7 \cr
& 13 & 173 & 612.8 \cr
& 14 & 666 & 3131.3 \cr
\noalign{\smallskip\hrule} \cr
}}$$
\end{table}

\clearpage
\pagebreak
\newpage

\end{document}